\documentstyle[12pt]{article}
\begin{document}

\begin{center}
{\Large
States and Observables in Semiclassical Field Theory: \\[0.2cm]
a Manifestly  Covariant  Approach
}
\\[1cm]
{{\large  O.Yu.Shvedov} \\[0.5cm]
{\it
Sub-Dept. of Quantum Statistics and Field Theory},\\
{\it Dept. of Physics, Moscow State University},\\
{\it 119992, Moscow, Vorobievy Gory, Russia}
}
\end{center}

\def\qp{
\mathrel{\mathop{\bf x}\limits^2},
\mathrel{\mathop{-i\frac{\partial}{\partial {\bf x}}}\limits^1} 
}
\def\gsim{{> \atop \sim}}
\def\lsim{{< \atop \sim}}

\setcounter{page}{0}

\begin{flushright}
hep-th/0412301
\end{flushright}

\section*{Abstract}

A manifestly covariant formulation of quantum field Maslov complex-WKB
theory (semiclassical field theory) is investigated for  the  case  of
scalar field.  The main object of the theory is "semiclassical bundle".
Its base is the set of all classical states, fibers are Hilbert spaces
of quantum  states in the external field.  Semiclassical Maslov states
may be viewed as points  or  surfaces  on  the  semiclassical  bundle.
Semiclassical analogs  of  QFT  axioms are formulated.  A relationship
between covariant   semiclassical   field   theory   and   Hamiltonian
formulation is discussed. The constructions of axiomatic field theory
(Schwinger sources, Bogoliubov $S$-matrix, Lehmann-Symanzik-Zimmermann
$R$-functions) are  used  in  constructing the covariant semiclassical
theory. A new covariant  formulation  of  classical  field  theory  and
semiclassical quantization proposal are discussed.

{\it Keywords:}
Maslov semiclassical   theory,   axiomatic   quantum   field   theory,
Bogoliubov S-matrix,  Lehmann-Symanzik-Zimmermann approach,  Schwinger
sources, Peierls brackets.

\footnotetext{e-mail:  shvedov@qs.phys.msu.su}

\footnotetext{This work was supported by the Russian  Foundation  for
Basic Research, project 02-01-01062}

\makeatletter
\@addtoreset{equation}{section}
\makeatother
\renewcommand{\theequation}{\thesection.\arabic{equation}}

\def\lb#1{\label{#1}}
\def\l#1{\lb{#1}}
\def\r#1{(\ref{#1})}
\def\c#1{\cite{#1}}
\def\i#1{\bibitem{#1}}
\def\beq{\begin{equation}}
\def\eeq{\end{equation}}
\def\bez{\begin{displaymath}}
\def\eez{\end{displaymath}}
\def\beb#1\l#2\eeb{\begin{equation} \begin{array}{c} #1 \qquad
\end{array} \label#2  \end{equation}}
\def\bey#1\eey{\begin{displaymath}
\begin{array}{c} #1  \end{array}  \end{displaymath}}

\newpage

\section{Introduction}

Semiclassical methods are very important for quantum mechanics and
field theory.  The  most convenient and general semiclassical approach
for quantum mechanics is the Maslov  theory  of  Lagrangian  manifolds
with complex germs \c{1,2}.  It is based on the direct substitution of
the hypothetical wave function to the Schrodinger equation, possesses the
mathematically rigorous justification (estimation of the accuracy) and
involves other semiclassical methods as partial cases:  the  Ehrenfest
approach,  the  oscillator  approximation,  the  wave  packet  and WKB
methods, the Maslov canonical operator method \c{3}.

When one generalizes the Maslov complex germ  theory  to  the  quantum
field models,  one  should  take into account the specific features of
QFT such  as  relativistic  Poincare   invariance,   divergences   and
renormalization. Direct application of the Maslov approach is possible
in Hamiltonian formulation of QFT only: one should first construct the
Hamiltonian of  the theory from the Lagrangian,  write the Schrodinger
equation with the help of the canonical quantization technique and use
the Maslov substitution. It is not easy to investigate divergences and
Poincare invariance in within this framework.

Some results  of  the  Hamiltonian  semiclassical  field  theory   are
presented in section 2. The QFT system with the Lagrangian depending on
the small parameter $h$ ("Planck constant") as
\beq
{\cal L} = \frac{1}{2} \partial_{\mu} \varphi \partial^{\mu} \varphi -
\frac{1}{h} V(\sqrt{h}\varphi),
\l{1.1}
\eeq
is considered.  Here $V(\Phi) \sim \frac{m^2}{2} \Phi^2$  as  $\Phi\to
0$. Then  it  is  possible  to  write  the  corresponding  Schrodinger
equation. The Maslov complex-WKB (complex-germ) substitution is
\beq
\Psi^t =
e^{\frac{i}{h} \tilde{S}^t}
e^{\frac{i}{\sqrt{h}} \int d{\bf x} [\Pi^t({\bf x}) \hat{\varphi}({\bf
x}) - \Phi^t({\bf x}) \hat{\pi}({\bf x})]} f^t
\l{1.2}
\eeq
One can  show  that  under  certain  conditions  the  ansatz   \r{1.2}
satisfies the   Schrodinger   equation   as  $h\to  0$;  relations  on
$\tilde{S},\Pi,\Phi,f$ can be written; in particular, $\Pi$ and $\Phi$
should obey classical field equations.

In addition  to  semiclassical evolution,  it is possible to construct
semiclassical Poincare  transformations,  field  operators  and  other
structures.

Contrary to  the  Hamiltonian  approach,  an  axiomatic  quantum field
theory \c{4} is based on the  most  general  principles,  such  as  Poincare
invariance, unitarity       and      causality.      Examples      are
Lehmann-Symanzik-Zimmermann (LSZ) \c{5}, Bogoliubov \c{6}
and Bogoliubov-Medvedev-Polivanov \c{7} theories.
The axiomatic perturbation theory is manifestly covariant.
This simplifies an analysis of divergences and  renormalization in
high orders of perturbation theory \c{6,8,9}.

Subsection 3.1 deals with semiclassical approximation in the axiomatic
theory. Instead of states \r{1.2} (with non-covariant decomposition of
space and time) one investigates "regularized" semiclassical states
of the form
\beq
\Psi = e^{\frac{i}{h}\overline{S}}
Texp [\frac{i}{\sqrt{h}} \int dx J(x) \hat{\varphi}_h(x)] \overline{f}
\equiv e^{\frac{i}{h}\overline{S}} T^h_J \overline{f},
\l{1.3}
\eeq
Here $\hat{\varphi}_h(x)$ is a Heisenberg field operator;
$J(x)$ is a c-number function (Schwinger classical source, cf.\c{10}).
It is supposed to vanish, except for the case
$x^0 \in [T_-,T_+]$. $Texp$ is a T-exponent.

It happens  that  the  following  property is satisfied in the leading
order in $h$ as $t>T_+$:
\beq
\hat{\varphi}_h (x) T^h_J \overline{f} \simeq
\frac{1}{\sqrt{h}} \overline{\Phi}(x) T^h_J \overline{f},
\l{1.4}
\eeq
Here the function $\overline{\Phi}(x) \equiv \overline{\Phi}_J(x)$
is a solution of the Cauchy problem
\beq
\partial_{\mu} \partial^{\mu}  \overline{\Phi} + V'(\overline{\Phi}) =
J, \qquad \overline{\Phi}|_{t<T_-} = 0.
\l{1.5}
\eeq
It is a classical field generated by the source $J$.

It is shown in section 4 that the state
\r{1.3} approximately equals to \r{1.2} under certain conditions on
$(\overline{S},J,\overline{f})$ and
$(\tilde{S},\Pi,\Phi,f)$. For the important partial case
$T_+<0$, the functions $\Phi$ and $\Pi$  coincides with classical field
$\overline{\Phi}$  and its time derivative as $t=0$ correspondingly.

There is the following specific feature of the covariant  approach  to
semiclassical field theory. Different sources
$J_1 \ne J_2$ may generate the same semiclassical state
\r{1.3},  provided that the fields
$\overline{\Phi}_{J_1}$ and  $\overline{\Phi}_{J_2}$ generated by them
coincide at  $t>T_+$.  Therefore,  one  should  take  into account the
equivalence relation between semiclassical states analogously to gauge
invariance in   gauge   theories.   This  property  is  considered  in
subsection 3.2. In particular, for the fields
$\overline{\Phi}_J$ with compact support, one has
\beq
T^h_J \overline{f} \simeq e^{\frac{i}{h} \overline{I}} W \overline{f}
\l{1.6}
\eeq
for some phase $\overline{I}$ and operator $W$ being an $S$-matrix  in
the classical background field $\overline{\Phi}$.

Starting from  relation  \r{1.6},  one  can derive {\it from the first
principles  }  classical  equations  of  motion   (stationary   action
principle)  and  commutation  relations  for semiclassical fields {\it
without using the postulate of canonical quantization}. Therefore, one
can   investigate   a   relationship   between  axiomatic  theory  and
Hamiltonian formalism.  For operator $W_J$,  one  obtains  analogs  of
Bogoliubov axioms of Poincare invariance, unitarity, causality.

\section{Hamiltonian semiclassical field theory}

{\bf 2.1.}
Consider the  Schrodinger  equation  for  the  model  with  Lagrangian
\r{1.1}:
\beq
i\frac{d}{dt} \Psi(t) = \int d{\bf x} \left[
\frac{1}{2} \hat{\pi}^2({\bf x}) +
\frac{1}{2} (\nabla \hat{\varphi}({\bf x}))^2
+ \frac{1}{h} V(\sqrt{h} \hat{\varphi}({\bf x}))
\right] \Psi(t).
\l{2.2}
\eeq
Here $\Psi(t)$ is a time-dependent state vector (element of
the space ${\cal H}^h$),  $\hat{\varphi}$ and $\hat{\pi}$ are
field and momentum operators.  They satisfy the well-known  canonical
commutation relations.   In   functional  Schrodinger  representation,
states at fixed time moment are present as functionals
$\Psi[\varphi(\cdot)]$, while operators
$\hat{\varphi}$ and $\hat{\pi}$ are written as
$\hat{\varphi}({\bf x}) \equiv \varphi({\bf x})$,
$\hat{\pi}({\bf x}) \equiv - i \frac{\delta}{\delta \varphi({\bf x})}$.
Semiclassical state \r{1.2} is presented as
\beq
\Psi[t,\varphi(\cdot)] =
e^{\frac{i}{h}S^t}
e^{\frac{i}{h} \int d{\bf x} \Pi^t({\bf x}) [\varphi({\bf x}) \sqrt{h}
- \Phi^t({\bf        x})]        }        f^t[\varphi(\cdot)         -
\frac{\Phi^t(\cdot)}{\sqrt{h}}] \equiv
(K^h_{S^t,\Pi^t,\Phi^t} f^t) [\varphi(\cdot)],
\l{2.4}
\eeq
Here $S^t  =  \tilde{S}^t  +  \frac{1}{2}\int  d{\bf  x} \Pi^t({\bf x})
\Phi^t({\bf x})$,  $f^t$  is an $h$-independent functional.
Substituting functional \r{2.4}  to the Schrodinger
equation, one obtains the following relations \c{11,12}:
\\
(а) $\Pi,\Phi$ obey the classical equations:
\beq
\dot{\Phi}^t = \Pi^t, \qquad -\dot{\Pi}^t = - \Delta \Phi^t + V'(\Phi^t);
\l{2.5}
\eeq
(б) $S$ is a classical action, i.e.
\beq
\dot{S}^t = \int d{\bf x}
[\Pi^t({\bf x}) \dot{\Phi}^t({\bf x}) - \frac{1}{2} (\Pi^t({\bf x}))^2
- \frac{1}{2} (\nabla \Phi^t({\bf x}))^2 -
V(\Phi^t({\bf x}))];
\l{2.6}
\eeq
(в) $f^t$   satisfies the Schrodinger equation in
the external field $\Phi^t({\bf x})$ with a quadratic Hamiltonian
\beb
i \dot{f}^t[\phi(\cdot)] = H_2^t
f^t[\phi(\cdot)]. \\
H_2^t =
\int d{\bf x}
\left[
- \frac{1}{2} \frac{\delta^2}{\delta \phi({\bf x}) \delta \phi({\bf x})}
+ \frac{1}{2} (\nabla \phi({\bf x}))^2 +
\frac{1}{2} V'{}'(\Phi^t({\bf x})) \phi^2({\bf x})
\right]
\l{2.7}
\eeb

To eliminate quantum field divergences in the  leading  order  of  the
semiclassical expansion,   one  should  choose  (generally,  non-Fock)
representation of the canonical commutation relations in the  external
field and add to the quadratic Hamiltonian an infinite
c-number $\Phi$-dependent counterterm \c{12,13}.

{\bf 2.2.} Thus, semiclassical states can be identified with pairs
$(X,f)$, where $X\equiv (S,\Pi,\Phi)$ are classical variables at fixed
moment of time,  $f$ (quantum state  in  the  external  field)  is  an
element of  an  $X$-dependent Hilbert space.  Evolution is viewed as a
set of transformations $u_t:  X  \mapsto  u_tX$,
$U_t(u_tX \gets  X):  {\cal  F}_X \to {\cal F}_{u_tX}$.
From the mathematical point of view, set of pairs
$(X,f)$  can be interpreted \c{14} as a vector bundle with a base
$\{ X\}$ and fibers ${\cal F}_X$ ("semiclassical  bundle").  Evolution
may be   viewed   as   a   1-parametric   automorphism  group  of  the
semiclassical bundle. The operator
$K_X^h :  {\cal F}_X \to {\cal H}^h$
taking a semiclassical state to quantum state \r{2.4} is called  as
{\it a canonical operator}.

Let ${\cal  U}_t^h$ be evolution transformation in the "exact" theory.
Then one writes:
\beq
{\cal U}^h_t K_X^h f \simeq K^h_{u_tX} U_t(u_tX \gets X) f.
\l{2.1a}
\eeq
Relation \r{2.1a} can be generalized to the case of
transformation ${\cal U}_g^h$
corresponding to an element of the Poincare
group $g\in G$:
\beq
{\cal U}_g^h K_X^f \simeq K_{u_gX}^h U_g(u_gX \gets X) f.
\l{2.8a}
\eeq
where
\beq
u_g: X  \mapsto u_gX,  \qquad U_g(u_gX \gets X):  {\cal F}_X \to {\cal
F}_{u_gX}.
\l{2.8}
\eeq
Construction of renormalized transformations \r{2.8}
and check of the group property  are  nontrivial  in  the  Hamiltonian
approach \c{13}.

Formulate analogs  of  QFT  axioms  for  the Hamiltonian semiclassical
field theory.

{\bf A1.} {\it A semiclassical bundle is defined.  The  space  of  the
bundle is a set of states of the semiclassical theory. The base
$\cal X$ of the bundle is a set of classical states, the fibers
${\cal F}_X$  are  Hilbert  state  spaces  of  a quantum system in the
external field $X\in {\cal X}$.
}

{\bf А2 (relativistic invariance)}.  {\it
The Poincare group $G$ acts on the semiclassical bundle: to each
$g\in G$  an  automorphism  \r{2.8}  of  the  semiclassical  bundle is
assigned; the group property is satisfied:
\bey
u_{g_1g_2} = u_{g_1}u_{g_2}, \\
U_{g_1g_2}(u_{g_1g_2}X \gets X) =
U_{g_1}(u_{g_1g_2}X \gets u_{g_2} X)
U_{g_2}(u_{g_2}X \gets X).
\eey
}

Investigate an analog of the field axiom. Since at $t=0$
\bez
\hat{\varphi}({\bf x}) K_X^h f = K^h_X
\left[ \frac{\Phi({\bf x})}{\sqrt{h}} + \hat{\phi}({\bf x}) \right]
f,
\eez
where $\hat{\phi}({\bf  x})$ is a multiplicator by
$\phi({\bf x})$ in the functional Schrodinger representation,  for the
Heisenberg field operator one has:
\beb
\hat{\varphi}_h({\bf x},t) K_X^h f =
{\cal U}^h_{-t} \hat{\varphi}({\bf x}) {\cal U}^h_t K_X^h f \simeq
{\cal U}^h_{-t} \hat{\varphi}({\bf x}) K^h_{u_tX} U_t(u_tX \gets X) f =
\\
{\cal U}^h_{-t} K^h_{u_tX}
\left[ \frac{\Phi^t({\bf x})}{\sqrt{h}} + \hat{\phi}({\bf x})
U_t(u_tX \gets X) \right] f =
\\
K^h_X
\left[
\frac{\Phi^t({\bf x})}{\sqrt{h}} +
U_{-t}(X \gets u_t X)
\hat{\phi}({\bf x})
U_t(u_tX \gets X)
\right]
f.
\l{2.9c}
\eeb
Therefore, one has
\beq
\hat{\varphi}_h(x) K_X^h f \simeq
K_X^h
\left[ \frac{\Phi(x|X)}{\sqrt{h}} + \hat{\phi}(x|X) \right] f,
\l{2.9}
\eeq
Here $\Phi(x|X)$ is a c-number function. It depends on
the space-time argument $x$ and classical state $X$.
$\hat{\phi}(x|X)$ is an operator-valued distribution
in ${\cal F}_X$. The following equations are satisfied:
\bez
\partial_{\mu} \partial^{\mu} \Phi(x|X) + V'(\Phi(x|X)) = 0,
\qquad
\partial_{\mu} \partial^{\mu} \hat{\phi}(x|X) +
V''(\Phi(x|X)) \hat{\phi}(x|X) = 0.
\eez
They can be obtained from the Heisenberg equations of motion for
$\hat{\varphi}_h(x)$.

Consider the   property  of  Poincare  invariance  of  fields  in  the
semiclassical approximation.  In  the  "exact"  theory,  for  Poincare
transformation
$g=(a,\Lambda)$ of the form  $x'=\Lambda  x+a$, one has
\beq
{\cal U}^h_{g^{-1}} \hat{\varphi}_h(x) {\cal U}^h_g =  \hat{\varphi}_h(w_g
x), \qquad w_gx = \Lambda^{-1}(x-a).
\l{2.9n}
\eeq
Analogously to eq.\r{2.9c}, one obtains that
\bez
\hat{\varphi}_h(w_g x) K^h_X f \simeq
K_X^h
\left[
\frac{\Phi(w_gx|X)}{\sqrt{h}} + \hat{\phi}(w_gx|X)
\right] f.
\eez

One comes to the following semiclassical analogs of field axioms.

{\bf А3 (field operator)}.  {\it To each semiclassical state
$X \in {\cal X}$  a c-number function  $\Phi(x|X)$
and an operator-valued distribution
$\hat{\phi}(x|X)$ in ${\cal F}_X$ are assigned.}

{\bf A4   (relativistic covariance of fields)}.  {\it
The following properties are satisfied:
\beb
\Phi(w_gx|X) = \Phi(x|u_gX); \\
\hat{\phi}(w_gx|X) = U_{g^{-1}}(X \gets u_gX) \hat{\phi} (x|u_gX)
U_g(u_gX \gets X).
\l{2.9f}
\eeb
}

{\bf 2.3.}
One can also introduce other structures \c{14}
on the semiclassical  bundle.
They are  related  with semiclassical approximation rather than fields
or Poincare invariance.

Let us shift a classical state
$X \in {\cal X}$ by a small quantity
$h\delta X$ of the order $h$. Then the quantum state
$K_X^hf$ will be multiplied by a c-number. Denote it as
$e^{-i\omega_X[\delta X]}$:
\beq
K^h_{X+\delta X} f \simeq e^{-i\omega_X[\delta X]} K_X^h f.
\l{2.9y}
\eeq
If one shifts a classical state by a quantity of the order
$\sqrt{h}$, transformation  of  the  canonical  operator  becomes more
complicated:
\beq
K^h_{X+\sqrt{h}\delta X} \simeq const K^h_X e^{i\Omega_X[\delta X]} f.
\l{2.9a}
\eeq
Here $const$ is a c-number multiplicator,
$\Omega_X[\delta X]$ is a Hermitian operator.
Moreover, $\omega_X$   and   $\Omega_X$   are
c-number and operator-valued 1-forms on
$\cal X$ correspondingly.
It happens that under some general requirements
the commutator of operators
$\Omega_X[\delta   X]$   is a c-number.
The c-number  is  related with 2-form $d\omega_X$ (differential of the
1-form $\omega_X$):
\beq
[\Omega_X[\delta X];   \Omega_X[\delta   X']]  =  -i  d\omega_X(\delta
X,\delta X').
\l{2.10}
\eeq
For our case,
\beq
\omega_X[\delta X] = \int d{\bf x} \Pi({\bf x}) \delta \Phi({\bf x}) -
\delta S,
\l{2.11}
\eeq
\beq
\Omega_X[\delta X]   =   \int   d{\bf   x}   [\delta    \Pi({\bf    x})
\hat{\varphi}({\bf x}) - \delta \Phi({\bf x}) \hat{\pi}({\bf x})].
\l{2.12}
\eeq

An important property of 1-forms
$\omega$ and    $\Omega$    is    their   independence   of   symmetry
transformations. This property means that quantum state vectors
$K^h_{X^t+h\delta  X^t}  f^t$  and $K^h_{X^t+\sqrt{h}\delta X^t} f^t$
should approximately satisfy the Schrodinger equation, provided that
$\delta X^t$ is a solution to the variation system.
One comes to the following axioms.

{\bf А5}.{\it
The 1-forms
$\omega$   ($\omega_X[\delta X] \in {\bf  R}$)  and  $\Omega$  (
$\omega_X[\delta X]$ is an operator in ${\cal  F}_X$) on $\cal X$
are given.  The commutation relation \r{2.10} is satisfied.
}

{\bf А6 (relativistic invariance of 1-forms)}.{\it
Let $u_g(X+\delta X) \simeq X' + \delta X'$. Then
\beq
\omega_{X'}[\delta X'] = \omega_X[\delta X].
\l{2.12a}
\eeq
\beq
\Omega_{X'}[\delta X']   U_g(X'   \gets   X)    =    U_g(X'\gets    X)
\Omega_X [\delta X].
\l{2.12b}
\eeq
}

{\bf 2.4.}
The Maslov  theory  of  Lagrangian  manifolds  of complex germ \c{1,2}
allows us to construct  semiclassical  solutions  of  the  Schrodinger
equations that  differs from \r{2.4}.  One of possible formulations is
as follows \c{11,14,15}. One considers the states being superpositions
of \r{2.4}:
\beq
\int d\alpha K^h_{X(\alpha)} f(\alpha).
\l{2.17}
\eeq
Here $\alpha = (\alpha_1,...,\alpha_k)$,  $(X(\alpha),f(\alpha))$ is a
$k$-dimensional surface  in  a  space of the semiclassical bundle.  It
happens that state \r{2.17} is of an exponentially small as
$h\to 0$ norm, except for the special case
\beq
\omega_X[\frac{\partial X}{\partial \alpha_a}] = 0.
\l{2.18}
\eeq
It is the only case to be considered. Under condition
\r{2.18} ("the Maslov isotropic condition")
the square of the norm of \r{2.17} is
\beq
h^{k/2} \int       d\alpha       (f(\alpha),       \prod_a        2\pi
\delta(\Omega_X[\frac{\partial X}{\partial \alpha_a}]) f(\alpha)).
\l{2.19}
\eeq
Therefore, for normalization it is necessary to multiply state
\r{2.17}  by $h^{-k/4}$.
The delta-functions entering to \r{2.19} commute because of relations
\r{2.10} and \r{2.18}.

Note that  formula \r{2.19} resembles the inner product that arises in
one of the approaches to quantize the constrained systems
\c{16}.

Under Poincare transformations,
state \r{2.17} is taken to
\bez
h^{-k/4} \int d\alpha K^h_{X(\alpha)} f(\alpha) \mapsto
h^{-k/4} \int d\alpha K^h_{u_gX(\alpha)}
U_g(u_gX(\alpha) \gets X(\alpha)) f(\alpha).
\eez
Invariance of 1-forms $\omega$ and $\Omega$ is a main requirement of
self-consistence of  the  Maslov  theory  of Lagrangian manifolds with
complex germs since the Maslov isotropic condition \r{2.18}, as well
as the   inner   product   \r{2.19}  should  conserve  under  Poincare
transformations.

{\bf 2.5.}
The covariant   formulation   of  semiclassical  field  theory  to  be
considered below resembles the semiclassical mechanics of  constrained
systems \c{17}.  A  specific features of both theories is that some of
semiclassical states  are  equivalent.  Namely,  for   any   pair   of
classically equivalent  states  $X\sim  X'$  an  isomorphism of fibers
$V(X'\gets X):  {\cal  F}_X  \to  {\cal  F}_{X'}$  is  specified.  The
following property should be satisfied:
\beq
V(X'' \gets X) = X(X'' \gets X') V(X'\gets X).
\l{2.20}
\eeq
Semiclassical states
$(X,f) \sim (X',V(X'\gets X)f)$
are called gauge-equivalent.

Semiclassical Poincare transformations, 1-forms $\omega$ and $\Omega$,
functions   $\Phi$   and   $\hat{\phi}$   should  conserve  the  gauge
equivalence relation. Therefore,
\beq
\begin{array}{c}
u_gX \sim u_gX', \\
V(u_gX'\gets u_gX) U_g(u_gX \gets X) =
U_g(u_gX' \gets X') V(X'\gets X);
\end{array}
\mbox{ as } X\sim X'
\l{2.21}
\eeq
\beq
\begin{array}{c}
\Phi(x|X) = \Phi(x|X'), \\
V(X'\gets X) \hat{\phi}(x|X) =
\hat{\phi}(x|X') V(X'\gets X)
\end{array}
\mbox{ as } X\sim X'
\l{2.22}
\eeq
\beq
\begin{array}{c}
\omega_X[\delta X] = \omega_{X'}[\delta X'], \\
V(X'\gets X) \Omega_{X}[\delta X] =
\Omega_{X'}[\delta X'] V(X'\gets X).
\end{array}
\mbox{ as }
\begin{array}{c}
X\sim X', \\ X+\delta X \sim X'+\delta X'.
\end{array}
\l{2.23}
\eeq
In particular, under infinitesimal gauge transformation
$X+\delta X \sim X$ the relations
$\omega_X[\delta X] = 0$, $\Omega_X[\delta X] = 0$ should be satisfied.
Requirements \r{2.20},  \r{2.21},  \r{2.22}, \r{2.23} are related with
self-consistence of semiclassical theory.

\section{An axiomatic approach to semiclassical field theory}

\subsection{Semiclassical states}

Consider an  axiomatic  approach  to construct the semiclassical field
theory. We will not  use  equations  of  motion;  instead,  the  usual
requirements \c{4}  of  the  axiomatic theory will be used.  These are
Poincare invariance,  existence of vacuum  state  $|0>$,  relativistic
invariance of the field
$\hat{\varphi}_h(x)$ and T-exponent
\beq
T_J^h = Texp\{\frac{i}{\sqrt{h}} \int dx J(x) \hat{\varphi}_h(x) \}.
\l{3.1}
\eeq
Other requirements will be also written explicitly.

{\bf Requirement  1.}  {\it  Hilbert spaces ${\cal H}^h$
coincide, while operators $\hat{\varphi}_h(x)$ and ${\cal
U}_g^h$ are expanded into asymptotic series in $\sqrt{h}$.
}

Requirement 1 is usual,  for example,  for the $S$-matrix approach  in
the asymptotic in-representation
\c{4,8}.

Denote these   operators   in   the   leading   order   in   $h$    as
$\hat{\varphi}_0(x)$ and $U_g$,  while the corresponding Hilbert space
will be denoted as ${\cal H}^0$.

Consider the semiclassical states of the form \r{1.3}:
\beq
\Psi =    e^{\frac{i}{h}\overline{S}}    T_J^h   \overline{f}   \equiv
\overline{K}_{\overline{S},J}^h \overline{f},
\l{3.2}
\eeq
Here $J(x)$  is a function with a compact support
in the domain $x^0 \in [T_-,T_+]$.
Identify states \r{3.2} with points on the semiclassical  bundle.  The
base is a set of pairs
$\overline{X}  =  \{
\overline{S},J(x)\}$. The fibers are Hilbert spaces
${\cal H}$.  Investigate the  axioms  of  semiclassical  field  theory
formulated in the previous section.

{\bf А2:} Since ${\cal  U}_g^h  T^h_J  \overline{f}  =  T^h_{u_gJ}
{\cal U}^h_g \overline{f} \simeq T^h_{u_gJ} U_g \overline{f}$, где $u_gJ(x) =
J(w_gx)$, $w_g$ has the form \r{2.9n},
the Poincare group acts on the semiclassical bundle as
follows: $(\overline{S},J,\overline{f})
\mapsto (\overline{S},u_gJ,U_g\overline{f})$.

{\bf А3:} Investigate the field operator as
$x^0>T_+$. One can write
$\hat{\varphi}(x) T^h_J = \frac{\sqrt{h}}{i} \frac{\delta T^h_J}
{\delta J(x)}$ and
\beq
\hat{\varphi}_h(x) T^h_J \overline{f} =
T^h_J \frac{1}{\sqrt{h}} R(x|J) \overline{f},
\l{3.3a}
\eeq
where
\beq
R(x|J) = - ih (T_J^h)^+ \frac{\delta T^h_J}{\delta J(x)}
\l{3.3}
\eeq
is a well-known LSZ $R$-function
\c{4,5}.   Impose the following requirement on it.

{\bf Requirement  2.} {\it $R$-function is expanded into an asymptotic
series in $\sqrt{h}$:
\bez
R(x|J) = \overline{\Phi}(x|J) + \sqrt{h} R^{(1)}(x|J) + ...
\eez
The leading order
$\overline{\Phi}(x|J)$ is a c-number.
}

Under this requirement, the c-number function
$\Phi(x|\overline{X})$  and the operator-valued distribution
$\hat{\phi}(x|\overline{X})$ have the following form as
$x^0>T_+$
\beq
\Phi(x|\overline{X}) = \overline{\Phi}(x|J), \quad
\hat{\phi}(x|\overline{X}) = R^{(1)}(x|J), \quad x^0>T_+.
\l{3.4}
\eeq

{\bf А5:} 1-forms $\omega$ and $\Omega$ can be introduced
according to ref.\c{17a}:
\bez
ih\delta \{e^{\frac{i}{h}\overline{S}} T_J^h\} \simeq
T_J^h
\{\overline{\omega}_{\overline{X}}[\delta \overline{X}]
- \sqrt{h} \overline{\Omega}_{\overline{X}}[\delta \overline{X}]\};
\eez
therefore,
$\overline{\omega}_{\overline{X}}[\delta \overline{X}]
- \sqrt{h} \overline{\Omega}_{\overline{X}}[\delta \overline{X}]
\simeq - \int dx R(x|J) \delta J(x) - \delta \overline{S}$
and
\beb
\overline{\omega}_{\overline{X}}[\delta \overline{X}] =
- \int dx \overline{\Phi}(x|J) \delta J(x) - \delta \overline{S}; \\
\overline{\Omega}_{\overline{X}}[\delta \overline{X}] =
\int dx R^{(1)}(x|J) \delta J(x).
\l{3.5}
\eeb
Thus, the main objects of the semiclassical field theory are presented
via the LSZ $R$-function. The following properties of the $R$-function
are well-known
\c{4}.

{\bf Proposition 1.} {\it
(а) $R(x|J)$  is a Hermitian operator;
it depends only on $J(y)$ at $y^0<x^0$ (Bogoliubov causality condition);
as $x^0<T_-$, it has the form
$R(x|J) = \hat{\varphi}_h(x) \sqrt{h}$.\\
(б) $R$-function is Poincare invariant:
${\cal    U}_{g^{-1}}^h R(x|u_gJ) {\cal U}_g^h = R(w_gx|J)$;\\
(в) the following commutation relation is satisfied:
\beq
[R(x|J);R(y|J)] = -ih
\left(
\frac{\delta R(x|J)}{\delta J(y)} -
\frac{\delta R(y|J)}{\delta J(x)}
\right).
\l{3.6}
\eeq
}

{\bf Corollary}. {\it 1. The function $\overline{\Phi}(x|J)$ is real,
vanishes as $x^0<T_-$,  depends only on $J(y)$ at $y^0 <
x^0$ and satisfies the Poincare invariance
property $\overline{\Phi}(w_gx|J) = \overline{\Phi}(x|u_gJ)$. \\
2. The operator distribution
$R^{(1)}(x|J)$ is Hermitian, coincides with
$\hat{\varphi}_0(x)$ при  $x^0<T_-$, depends only on $J(y)$ at
$y^0<x^0$, satisfies the relativistic invariance property
\beq
U_{g^{-1}} R^{(1)}(x|u_gJ) U_g = R^{(1)}(w_gx|J)
\l{3.6a}
\eeq
and commutation relation
\beq
[R^{(1)}(x|J);R^{(1)}(y|J)] = - i
\left(
\frac{\delta \overline{\Phi}(x|J)}{\delta J(y)} -
\frac{\delta \overline{\Phi}(y|J)}{\delta J(x)}
\right).
\l{3.7}
\eeq
}

Thus, relativistic invariance of fields and 1-forms is a corollary  of
general properties of the LSZ $R$-functions.  The commutation relation
\r{2.7} coincides with \r{2.10}.

\subsection{Equivalence of semiclassical states}

It has been written in section 1 that some semiclassical states may be
equivalent each other in the covariant approach. Say that
$J \sim 0$ iff
\beq
T^h_J \overline{f}   \simeq    e^{\frac{i}{h}    \overline{I}_J}    W_J
\overline{f}
\l{3.8}
\eeq
for some c-number
$\overline{I}_J$ and operator $W_J$ being an asymptotic series in
$\sqrt{h}$.

Denote by $W_J^0$  the operator $W_J$ in the leading order
in $\sqrt{h}$. Investigate the properties
of $\overline{I}_J$ and $W_J$.

{\bf Proposition 2.} {\it
1. The operator $W_J$ is unitary. \\
2. Let $J \sim 0$. Then $u_gJ \sim 0$ and
\beq
\overline{I}_{u_gJ} = \overline{I}_J, \qquad
U_g W_J^0 U_g^{-1} = W^0_{u_gJ}.
\l{3.9}
\eeq
3. Under condition $x \gsim supp J$ the following properties are
satisfied:
\beq
\overline{\Phi}(x|J) = 0, \quad
R^{(1)}(x|J) = (W_J^0)^+ \hat{\varphi}_0(x) W_J^0.
\l{3.10}
\eeq
}

The first property is a corollary of unitarity of the operator
$T_J^h$. The second one can be obtained by applying the operator
${\cal U}_g^h$ to the relation \r{3.8}.
The properties \r{3.10} are corollaries of the relation
$T^h_J R(x|J)   \overline{f}   =   \sqrt{h}   \hat{\varphi}_h(x)  T^h_J
\overline{f}, \qquad x \gsim supp J$. It is taken to the form
$R(x|J) =  \sqrt{h}  W_J^+  \hat{\varphi}(x)
W_J$ and expanded into a perturbation series.

{\bf Proposition 3.} {\it
Under conditions $J\sim 0$, $J+\delta J \sim 0$
\beq
\delta \overline{I} = \int dx \overline{\Phi}(x|J) \delta J(x),
\qquad \int dx R^{(1)}(x|J) \delta J(x) = 0.
\l{3.11}
\eeq
}

To check eq.\r{3.11}, consider the variation of relation
\r{3.8}:
\\ $\delta T_J^h   \cdot   \overline{f}    \simeq    \frac{i}{h}    \delta
\overline{I}_J T_J^h \overline{f} +
T_J^h W_J^+ \delta W_J \overline{f}$.
Therefore,
\beq
-ih (T_J^h)^+ \delta T_J^h = \delta \overline{I}_J - ih  W_J^+  \delta
W_J.
\l{3.12}
\eeq
On the other hand, it follows from eq.\r{3.3} that
\beq
-ih (T_J^h)^+ \delta T_J^h = \int dx R(x|J) \delta J(x).
\l{3.13}
\eeq
Comparing eqs.\r{3.12} and \r{3.13}, we obtain relation \r{3.11}.

The following  statement is a corollary of the causality principle for
$W_J^0$.

{\bf Proposition 4.} {\it
Let
$J + \Delta J_2 \sim 0$,  $J + \Delta J_1 + \Delta J_2 \sim  0$,
$supp \Delta J_2 \gsim supp \Delta J_1$.
Then the operator $(W_{J+\Delta
J_2}^0)^+ W^0_{J+\Delta      J_1+\Delta      J_2}$   and c-number
$-\overline{I}_{J+\Delta J_2}  +  \overline{I}_{J+\Delta  J_1 + \Delta
J_2}$ do not depend on $\Delta J_2$.
}

To check the proposition, it is sufficient to consider the operator
\\ $(T^h_{J+\Delta J_2})^+ T^h_{J+\Delta J_1+\Delta J_2} \simeq
e^{\frac{i}{h}
[-\overline{I}_{J+\Delta J_2}  +  \overline{I}_{J+\Delta  J_1 + \Delta
J_2}]} (W_{J+\Delta J_2}^0)^+ W^0_{J+\Delta J_1+\Delta J_2}
$. It is $\Delta J_2$-independent.

Introduce an  equivalence  relation  on the semiclassical bundle.  Say
that $J_1 \sim J_2$ iff the properties $J_1+J_+ \sim 0$,  $J_2  +  J_+
\sim  0$ are satisfied for some source $J_+$ with $supp J_+ \gsim supp
J_1 \cup supp J_2$

{\bf Proposition 5.} {\it Let $J_1 \sim J_2$. Let also the function
$J_+'$  satisfy the properties
$supp J_+' \gsim supp J_1 \cup supp J_2$ и $J_1 + J_+' \sim 0$.
Then $J_2 + J_+' \sim 0$.
}

To check the proposition, it is sufficient to use the property
\\$T^h_{J_2+J_+'} = T^h_{J_1+J_+'} (T^h_{J_2+J_+})^+ T^h_{J_1+J_+}$.

{\bf Corollary.}  {\it Let $J_1 \sim J_2$ and $J_2 \sim J_3$. Then
$J_1 \sim J_3$. }

{\bf Proposition 6.} {\it Let $J_1 \sim J_2$. Then
the relation
\beq
e^{\frac{i}{h}\overline{S}_1} T^h_{J_1} \overline{f}_1
\simeq
e^{\frac{i}{h}\overline{S}_2} T^h_{J_2} \overline{f}_2
\l{3.14}
\eeq
is satisfied iff
\beq
\overline{S}_1 + \overline{I}_{J_1+J_+} =
\overline{S}_2 + \overline{I}_{J_2+J_+},
\qquad
\overline{f}_2 = (W^0_{J_2+J_+})^+ W^0_{J_1+J_+} \overline{f}_1.
\l{3.15}
\eeq
}

To check the proposition, one should apply the operator
$T^h_{J_+}$  to sides of the relation \r{3.14}.

Say that
two semiclassical states are equivalent,
$(\overline{X}_1,\overline{f}_1) \sim
(\overline{X}_2,\overline{f}_2)$, iff
relation \r{3.15} is satisfied.
Under our notations
\beq
V(\overline{X}_2 \gets \overline{X}_1) =
(W^0_{J_2+J_+})^+ W^0_{J_1+J_+} \equiv V(J_2 \gets J_1).
\l{3.16}
\eeq
The causality property (proposition 4) implies that definition
\r{3.16} does not depend on choice of the source $J_+$.

Let us check properties \r{2.20}  -- \r{2.23}.
Relation \r{2.20} is evident.
Eq.\r{2.21}   is a corollary of Poincare invariance
(proposition 2). Property
\r{2.23} for $\omega$ can be checked as follows. Let
$(\overline{S}_1,J_1) \sim (\overline{S}_2,J_2)$,
$(\overline{S}_1 + \delta \overline{S}_1,J_1 + \delta J_1)
\sim (\overline{S}_2  +  \delta  \overline{S}_2,J_2  +  \delta  J_2)$.
According to eq.\r{3.15}, this means that
\bez
\delta \overline{S}_1 + \int dx \overline{\Phi}(x|J_1) \delta J_1 =
\delta \overline{S}_2 + \int dx \overline{\Phi}(x|J_2) \delta J_2,
\eez
or $\omega_{\overline{X}_1}[\delta \overline{X}_1]
\omega_{\overline{X}_2}[\delta \overline{X}_2]$.

Let us now apply the operator
$\hat{\varphi}_h(x)$ to equality \r{3.14}.
One finds that
$e^{\frac{i}{h} \overline{S}_1} T^h_{J_1} R(x|J_1) \overline{f}_1 =
e^{\frac{i}{h} \overline{S}_2} T^h_{J_2} R(x|J_2) \overline{f}_2$.
Therefore,
\bez
R(x|J_2) V(J_2 \gets J_1) = V(J_2\gets J_1) R(x|J_1).
\eez
Expand this relation into a perturbation series. For
$x \gsim supp
J_1 \cup supp J_2$, one finds:
\beq
\overline{\Phi}(x|J_1) = \overline{\Phi}(x|J_2), \quad
R^{(1)}(x|J_2) V(J_2 \gets J_1) = V(J_2 \gets J_1) R^{(1)}(x|J_1).
\l{3.17}
\eeq
Thus, relation
\r{2.22} is checked at $x\gsim supp J$
($\Phi$ and $\hat{\phi}$ have been defined at these $x$ only).
Properties \r{2.22} can be used
to extend the functions $\Phi$ and
$\hat{\phi}$ and define them for other values of $x$.
To do this, one should choose a
source
$J' \sim J$ such that $supp J \lsim x$ and set
\bez
\Phi(x|\overline{X}) \equiv \overline{\Phi}(x|J'); \quad
\hat{\phi}(x|\overline{X}) \equiv
V(J \gets J') R^{(1)}(x|J') V(J'\gets J).
\eez
Properties \r{2.22} are valid for the extensions of functions as well.

Check the relation \r{2.23} for $\Omega$.

{\bf Proposition 7.} {\it Let
$J_1 \sim J_2$, $J_1 + \delta J_1 \sim J_2 + \delta J_2$.
Then
\beq
\int dx R^{(1)}(x|J_2) \delta J_2 V(J_2 \gets J_1) =
V(J_2 \gets J_1) \int dx R^{(1)}(x|J_1) \delta J_1(x).
\l{3.18}
\eeq
}

To check the proposition, introduce the
following operator function
\beq
\tilde{R}(x|J) = T^h_J R(x|J) (T^h_J)^+.
\l{3.19}
\eeq
It depends only on
$J(y)$  at $y^0>x^0$, obeys the boundary condition
$\tilde{R}(x|J)  = \hat{\varphi}(x)
\sqrt{h}$ as  $x  \gsim  supp J$.  For the case $J \sim 0$,
the $\tilde{R}$-function has the form
$\tilde{R}(x|J) = W_J R(x|J) W_J^+$,
can be expand into an asymptotic series
\bez
\tilde{R}(x|J) = \tilde{\Phi}(x|J) + \sqrt{h}  \tilde{R}^{(1)}(x|J)  +
...,
\eez
with
\beq
\tilde{\Phi}(x|J) = \Phi(x|J), \qquad
\tilde{R}^{(1)}(x|J) = W_J^0 R^{(1)}(x|J) (W_J^0)^+.
\l{3.20}
\eeq
Let $J_1\sim J_2$ и $J_1 + \delta J_1 \sim J_2  +  \delta  J_2$.  This
means that
$J_1 + J_+ \sim 0$,  $J_2 + J_+ \sim 0$,
$J_1 +  \delta J_1 + J_+ + \delta J_+ \sim 0$,
$J_2 + \delta J_2 + J_+ + \delta J_+ \sim 0$ at $supp J_+  \cup  supp
\delta J_+ \gsim supp J_1 \cup supp J_2 \cup supp \delta J_1 \cup supp
\delta J_2$.  It follows from eqs. \r{3.11},  \r{3.16},  \r{3.20},
that relation \r{3.18} is equivalent to the following:
\beb
(W^0_{J_2+J_+})^+ \int  dx  \tilde{R}^{(1)}  (x|J_2+J_+) \delta J_+(x)
W^0_{J_1+J_+} = \\
(W^0_{J_2+J_+})^+ \int  dx  \tilde{R}^{(1)}  (x|J_1+J_+) \delta J_+(x)
W^0_{J_1+J_+},
\l{3.21}
\eeb
However, eq.\r{3.21} is a corollary of the causality condition for
$\tilde{R}^{(1)}$:
$\tilde{R}^{(1)}(x|J_1+J_+)
= \tilde{R}^{(1)}(x|J_2+J_+) = \tilde{R}^{(1)}(x|J_+)$.
Proposition is justified.

To obtain the classical field equations "from the first principles",
impose also the following requirement.

{\bf Requirement 3.  } {\it For any field configuration
$\Phi(x)$ with a compact support there exists a unique source
$J \sim  0$ (denote it as $J=J(x|\Phi)$) generating
the field configuration $\Phi$:
$\Phi(x) = \overline{\Phi}(x|J)$. It satisfies the locality
condition
\beq
\frac{\delta J(x|\Phi)}{\delta \Phi(y)} = 0, y\ne x.
\l{3.22a}
\eeq
}

Requrement 3 and eq.\r{3.11} imply the following statement.

{\bf Poposition 8.} {\it
1. The functional
$I[\Phi]  =  \overline{I}_{J_{\Phi}}  -  \int  dx  J(x)
\Phi(x)$ satisfies the locality
property $I[\Phi] = \int  dx  {\cal
L}(\Phi(x),\partial_{\mu}\Phi(x),...,
\partial_{\mu_1}...\partial{\mu_n} \Phi(x))$ and stationary action
principle
\beq
\frac{\delta I[\overline{\Phi}]}{\delta \Phi(x)} = - J(x).
\l{3.22}
\eeq
2. The operator-valued distribution
$R^{(1)}(x|J)$  satisfies the equation
\beq
\int dy   \frac{\delta^2   I[\overline{\Phi}]}{\delta  \Phi(x)  \delta
\Phi(y)} R^{(1)}(y|J) = 0.
\l{3.23}
\eeq
In particular, the field
$\hat{\varphi}_0(x)$ obeys the equation
\beq
\int dy  \left.
\frac{\delta^2   I[\overline{\Phi}]}{\delta  \Phi(x)  \delta
\Phi(y)}\right\vert_{\Phi=0} \hat{\varphi}_0(y) = 0.
\l{3.23a}
\eeq
3. Denote by $D^{ret}_{\overline{\Phi}}(x,y)$ the retarded
Green function of the problem
\bez
\int dy   \frac{\delta^2   I[\overline{\Phi}]}{\delta  \Phi(x)  \delta
\Phi(y)} \delta \Phi(y) = -\delta J(x), \qquad
\delta \Phi|_{x^0<T_-} = 0.
\eez
It is defined from the relation
$\delta    \Phi(x)    =    \int   dy
D_{\overline{\Phi}}^{ret}(x,y) \delta J(y)$. Then
\beq
[R^{(1)}(x|J);R^{(1)}(y|J)] = - i
(D_{\overline{\Phi}}^{ret}(x,y) - D_{\overline{\Phi}}^{ret}(y,x)).
\l{3.24}
\eeq
In particular,
\beq
[\hat{\varphi}_0(x);\hat{\varphi}_0(y)] = - i
(D_{0}^{ret}(x,y) - D_{0}^{ret}(y,x)).
\l{3.24a}
\eeq
}

The property \r{3.22} is another
form of eq.\r{3.11}, since
\\ $\delta I = \delta \overline{I} - \int dx (\delta J \overline{\Phi} + J
\delta \overline{\Phi}) = - \int dx J\delta \overline{\Phi}$.
The locality property for $I$ is a corollary of \r{3.22a}.
Moreover, let $\Phi$   and   $\Phi  +  \delta  \Phi$  be
field configurations with compact support.
Then for corresponding sources
$J$, $J+\delta J$ one has
$\delta J(y) =
-
\int dx   \frac{\delta^2   I[\overline{\Phi}]}{\delta  \Phi(x)  \delta
\Phi(y)} \delta \Phi(x)$.
Therefore, eq.\r{3.11} is rewritten as
$- \int dx dy R^{(1)}(y|J)
\frac{\delta^2   I[\overline{\Phi}]}{\delta  \Phi(x)  \delta
\Phi(y)} \delta \Phi(x) = 0$.
One obtains then eq.\r{3.23} and its
partial case \r{3.23a}.
Commutation relation
\r{3.24} is another form of \r{3.7}.
Proposition is justified.

Thus, the  semiclassical  theory is reconstructed in the leading order
in $h$ from the action functional
$I[\Phi]$ {\it without postulates of the canonical quantization}.
Namely, starting from $I[\Phi]$, one uniquely finds:

- equation of motion \r{3.22}  for  $\Phi$;
it allows us to find $\overline{\Phi}(x|J)$ from the boundary condition
$\Phi|_{x^0<T_-}  =  0$;

- equation of motion \r{3.23a} and commutation relation \r{3.24a}  for
free field $\hat{\varphi}_0(x)$;

- equation of motion \r{3.23} for  the  operators  $R^{(1)}(x|J)$  and
commutation relation \r{3.24} for these operators.

The right-hand side of the obtained  relation  \r{3.24}  contains  the
Peierls  bracket  \c{19}  (see  also \c{20,21}) of classical fields at
points $x$ and $y$.  The Peierls bracket postulate was  considered  in
\c{20} as a foundation of QFT.

If one fixes the representation for the operators
$\hat{\varphi}_0$ (for example, the Fock representation) then

- relation $U_{g^{-1}}    \hat{\varphi}_0(x)    U_g    =
\hat{\varphi}_0(w_gx)$ allows us to reconstruct
the operator $U_g$ up to a c-number multiplier for the free theory;
the multiplier is reconstructed fro, the condition
$U_g|0> = |0>$;

- eq.\r{3.23}   and   initial    condition    $R^{(1)}|_{x^0<T_-}    =
\hat{\varphi}_0(x)$   allows   us   to   reconstruct   the   operators
$R^{(1)}(x|J)$ uniquely;

- relation \r{3.10} for $x^0>T_+$ allows us to reconstruct
the operator $W_J^0$ up to a c-number multiplier
$c_J$; it follows from the causality condition that
$\exp[i\int dx
{\cal L}_1(\Phi(x),\partial_{\mu}\Phi(x),...,
\partial_{\mu_1}...\partial{\mu_n} \Phi(x))]$,  here  ${\cal  L}_1$ is
a one-loop counterterm.

\section{Correspondence of Hamiltonian and axiomatic approaches }

{\bf 4.1.} We have considered Hamiltonian and  axiomatic  formulations
of semiclassical field theory. Investigate their correspondence.

First of all, show that state
\r{1.3} of the covariant approach can be taken to the form
\r{2.4}.  Notice that the operator
$T_J^h = Texp[\frac{i}{\sqrt{h}} \int dx J(x) \hat{\varphi}_h(x)]$
is related with the
evolution transformation $U_J(0,t_-)$ for the system with
Hamiltonian
\bez
H_J(t) =   H   -   \frac{1}{\sqrt{h}}   \int  d{\bf  x}  J({\bf  x},t)
\hat{\varphi}({\bf x})
\eez
via the standard relation
\bez
T_J^h = U_J(0,t_-) e^{-iHt_-}, \qquad t_- < (supp J)^0.
\eez
One also has
\beq
e^{-iHt_-} \overline{f} \simeq e^{-iH_0t_-} \overline{f},
\l{4.1}
\eeq
where $H_0$  is a Hamiltonian
of the free field with mass
$m$.
The semiclassical  state   \r{4.1}   is   taken   to   \r{2.4}   under
semiclassical evolution with Hamiltonian $H_J$;
one obtains the system of equations for
$S,\Pi,\Phi,f$; it is related to
\r{2.5},
\r{2.6}, \r{2.7}  by substitution
$V(\Phi) \Rightarrow V(\Phi) - J\Phi$.
Denote by $U_2(0,t_-)$
the evolution operator for eq.\r{2.7}. Then
\bez
e^{\frac{i}{h}\overline{S}} T_J^h \overline{f} \simeq
K^h_{S,\Pi,\Phi} f, \qquad (supp J)^0 = [T_-,T_+] \subset (-\infty,0).
\eez
Here
\beb
\Pi({\bf x}) = \dot{\overline{\Phi}}(x|J), \quad
\Phi({\bf x})   =  \overline{\Phi}(x|J),  \quad  S  =  \overline{S}  +
I_-[\Phi], \quad x^0=0 \\
I_-[\overline{\Phi}] = \int_{x^0<0} dx
[\frac{1}{2} \partial_{\mu}       \overline{\Phi}       \partial^{\mu}
\overline{\Phi} - V(\overline{\Phi}) + J\overline{\Phi}], \\
f = {\cal V}_{\overline{X}} \overline{f}, \quad
{\cal V}_{\overline{X}} = U_2(0,t_-) e^{-iH_0t_-}, \quad t_-<T_-.
\l{4.4}
\eeb
Thus, the property
\r{1.4} is obtained.

{\bf 4.2.} Investigate the correspondence of classical field  theories
in Hamiltonian and axiomatic aproaches.
For the former formulation,  an extended phase space (the base of the
semiclassical bundle) consists of sets
$(S,\Pi({\bf x}),\Phi({\bf x}))$; the 1-form $\omega$ is
\beq
\omega_X[\delta X] = \int d{\bf x} \Pi({\bf x}) \delta \Phi({\bf x}) -
\delta S.
\l{4.5}
\eeq
For the latter formulation,
an extended phase space may be viewed as a surface on the space of sets
$\{\overline{X}    =
(S,J(x),\overline{\Phi}(x))\}$; the equation of surface has the
form \r{3.22}:
\beq
\partial_{\mu} \partial^{\mu}           \overline{\Phi}(x)           +
V'(\overline{\Phi}(x)) = J(x), \quad \Phi|_{x^0<T_-} = 0.
\l{4.6}
\eeq
The 1-form $\omega$ is written as
\beq
\overline{\omega}_X[\delta \overline{X}]     =     -      \int      dx
\overline{\Phi}(x) \delta J(x) - \delta \overline{S}.
\l{4.7}
\eeq
Relations \r{4.6} and \r{4.7} allows us
to interpret $J(x)$ and
$\overline{\Phi}(x)$ as momenta and coordinates correspondingly,
while relation \r{4.6} is a set of first-class constraints.

Equivalence of  Hamiltonian  and  covariant  formulations of classical
field theory is a corollary of relation
$\overline{\omega}_{\overline{X}}
[\delta \overline{X}]    =    \omega_X[\delta    X]$.
It can be checked by a direct calculation. One should use properties
\r{4.4} and relation
$\delta S  =  \delta  \overline{S}  +  \delta   I_-[\overline{\Phi}]$.
The symplectic 2-forms $d\omega$
also coincide:
\bez
\int d{\bf x}
[\delta \Pi_1({\bf x}) \delta \Phi_2({\bf x})
- \delta \Pi_2({\bf x}) \delta \Phi_1({\bf x})] =
\int dx
[\delta J_1({x}) \delta \Phi_2({x})
- \delta J_2({x}) \delta \Phi_1({x})].
\eez

Note that  Poincare  transformation generators have the following form
in the covariant formulation:
\beb
{\cal P}^{\mu} = \int dx J(x) \partial^{\mu} \overline{\Phi}(x),
\\
{\cal M}^{\mu\nu} = \int dx J(x)
(x^{\mu} \partial^{\nu} - x^{\nu} \partial^{\mu}) \overline{\Phi}(x).
\l{4.8}
\eeb
Making use of eq.\r{4.6},  one can take the generators to the form  of
the Hamiltonian field theory. For example,
${\cal P}^{\mu}$ is an integral of
$T^{\mu 0}$ over space.
Analogously to eq.\r{4.8},  it is possible to write generators of  any
transformation that conserve the action.  The generators coincide with
Noether integrals of motion.

{\bf 4.3.}
Investigate now    the   relationship   between   other   objects   of
semiclassical field theory in Hamiltonian  and  axiomatic  approaches:
semiclassical Poincare transformations
$U_g(u_gX\gets X)$, semiclassical fields
$\hat{\Phi}(x|X)$,  1-forms  $\Omega$.
Equivalence of the approaches means that
\beq
{\cal V}_{u_g\overline{X}} U_g \overline{f} =
U_g(u_gX \gets X) {\cal V}_{\overline{X}} \overline{f};
\l{4.9}
\eeq
\beq
{\cal V}_{\overline{X}} R^{(1)}(x|J) \overline{f} =
\hat{\phi}(x|X) {\cal V}_{\overline{X}} \overline{f}; \qquad
x \gsim supp J;
\l{4.10}
\eeq
\beq
{\cal V}_{\overline{X}}
\int dx R^{(1)}(x|J) \delta J(x) \overline{f} =
\int d{\bf x} [\delta \Pi({\bf x}) {\phi}({\bf x}) -
\delta \Phi({\bf x}) \frac{1}{i} \frac{\delta}{\delta \phi({\bf x})}]
{\cal V}_{\overline{X}} \overline{f};
\l{4.11}
\eeq
Relation \r{4.11}  can be checked as follows.
First of all, take into account relation
\r{4.6} and property
\beq
\partial_{\mu} \partial^{\mu}       \delta      \overline{\Phi}      +
V''(\overline{\Phi}) \delta \overline{\Phi} = \delta J,
\qquad
\delta \overline{\Phi}|_{x\lsim supp \delta J} = 0
\l{4.11a}
\eeq
Take the operator in the left-hand side to the form:
\beb
\int dx R^{(1)}(x|J) \delta J(x) = \int dx R^{(1)}(x|J)
(\partial_{\mu} \partial^{\mu} \delta \overline{\Phi}(x) +
V''(\overline{\Phi}) \delta \overline{\Phi}(x)) = \\
\int dx [ R^{(1)}(x|J)
\partial_{\mu} \partial^{\mu} \delta \overline{\Phi}(x)
- \delta \overline{\Phi}(x)
\partial_{\mu} \partial^{\mu} R^{(1)}(x|J)] = \\
\int d\sigma^{\mu}      [R^{(1)}(x|J)      \partial^{\mu}       \delta
\overline{\Phi}(x) - \delta \overline{\Phi}(x) \partial^{\mu} R^{(1)}(x|J)]
\l{4.12}
\eeb
Integration in  \r{4.12} is performed over and space-like surface such
that $x\gsim supp J$.

By $\delta \Phi_+(x)$ we denote the solution of the problem
\beq
[\partial_{\mu} \partial^{\mu}  +  V'{}'(\overline{\Phi}(x))]   \delta
\Phi_+ = 0, \qquad \delta \Phi_+|_{x^0>T_+} = \delta \overline{\Phi};
\l{4.12a}
\eeq
then formula \r{4.12} is taken to the form
\beq
\int dx R^{(1)}(x|J) \delta J(x) =
\int d\sigma^{\mu}      [R^{(1)}(x|J)      \partial^{\mu}       \delta
{\Phi}_+(x) - \delta {\Phi}_+(x) \partial^{\mu} R^{(1)}(x|J)],
\l{4.12b}
\eeq
or
\beq
\int dx R^{(1)}(x|J) \delta J(x) =
\int d\sigma^{\mu}      [\hat{\varphi}_0(x)
\partial^{\mu}       \delta
{\Phi}_+(x) - \delta {\Phi}_+(x) \partial^{\mu} \hat{\varphi}_0(x)].
\l{4.13}
\eeq
The space-like surface of integration is arbitrary in eq.\r{4.12b} and
satisifes the condition $x \lsim supp J$ in eq.\r{4.13}.

Notice that the operator
$A(t) =   \int_{x^0=t}   d{\bf   x}   [\dot{\delta  \Phi}_+({\bf  x},t)
\hat{\varphi}({\bf x}) - \delta \Phi_+({\bf x},t) \hat{\pi}({\bf x})]$
commutes with $i\frac{d}{dt} - H_2^t$
and takes the solutions of eq.\r{2.7} to the solutions.
Therefore, the left-hand side of eq.\r{4.11} is as follows:
\bez
{\cal V}_{\overline{X}} e^{iH_0t_-} A(t_-) e^{-iH_0t_-} \overline{f} =
U_2(0,t_-) A(t_-)   e^{-iH_0t_-}   \overline{f}   =   A(0)  U_2(0,t_-)
e^{-iH_0t_-} \overline{f} = A(0) {\cal V}_{\overline{X}} \overline{f}
\eez
It coincides with the right-hand side. Relation \r{4.11} is checked.
Its left-hand side can be presented as an integral
\r{4.12} over the surface $x^0=0$; therefore, for $x^0=0$
\bez
{\cal V}_{\overline{X}} R^{(1)}(x|J) =
\hat{\phi}(x) {\cal V}_{\overline{X}},
\qquad
{\cal V}_{\overline{X}} \partial_0 R^{(1)}(x|J) =
\partial_0 \hat{\phi}(x)  {\cal V}_{\overline{X}},
\eez
Since $\hat{\phi}(x)$ and $R^{(1)}(x|J)$ satisfy the same second-order
differential equation at  $x\gsim  supp  J$,  one  comes  to  relation
\r{4.10}.

Property \r{4.9}  can  be  obtained  from  relativistic  invariance of
fields up to a c-number multiplier.  Namely,  the operators  $U_g(u_gX
\gets  X)$  are  found  from  \r{2.9f}  up to a multiplier,  while the
operator ${\cal V}_{u_g\overline{X}} U_g {\cal V}_{\overline{X}}^{-1}$
satisifes \r{2.9f}. To investigate the c-number multiplier, one should
analyze renormalization of Poincare transformations in more details.

Notice also that formulas \r{4.10} and \r{4.11} allows  us  to  obtain
useful relations for semiclassical field theory
$\hat{\phi}(x|X)$ and 1-form $\Omega$. For the simplicity,
denote
\bez
\Omega\{\delta \Phi(\cdot)\} = \Omega(\delta \Phi,  \delta \dot{\Phi})
= \int  d  {\bf  x}  [\delta \dot{\Phi}({\bf x},0) \hat{\phi}({\bf x}) -
\delta \Phi({\bf x},0) \hat{\pi}({\bf x})],
\eez
iff $\delta \Phi$ is a solution of equation
\bez
\partial_{\mu}\partial^{\mu} \delta \Phi + V''({\Phi}(x)) \delta  \Phi
= 0.
\eez
It follows from
\r{4.10}, \r{4.11}, \r{4.12b} that
\beq
\int dx \hat{\phi}(x|X) \delta J(x) = \Omega\{\delta \Phi_+(\cdot)\}.
\l{4.15}
\eeq
Here $\delta \Phi_+$ is found from relations \r{4.11a} and \r{4.12a}.

Property \r{4.15} shows us that axioms of semiclassical field and  its
Poincare invariance  are  not  independent.  It  is possible to define
semiclassical field $\hat{\phi}(x|X)$  by  eq.\r{4.15}.  Its  Poincare
invariance is a corollary of relativistic invariance of 1-form
$\Omega$.

\section{Conclusions}

Thus, it  is  possible  to  derive  main  formulas  of  classical  and
semiclassical field theroy (stationary action proinciple, equations of
motion and commutation relations for semiclassical  fields)  from  the
first principles of axiomatic QFT and general requiremnets.

The main  difficulty  of axiomatic QFT is that there are no nontrivial
model obeying  all  the  axioms.  Therefore,  when  one   investigates
models of semiclassical field theory, it seems to be reasonable to
consider the properties that are essential in the leading order of $h$
and check  only  them.  These essential properties are axioms A1-A6 in
Hamiltonian approach.  For  the   covariant   approach,   one   should
additionally investigate properties \r{2.20} -- \r{2.23}.

To study semiclassical perturbation theory,  one should generalize the
considered axioms.  It is also interesting to construct  semiclassical
gauge theories  in  the  axiomatic  approach.  The authors is going to
consider these problems in the following publications.

\end{document}